\documentclass[pra,twocolumn,preprintnumbers,amsmath,amssymb,superscriptaddress]{revtex4}

\usepackage{color}
\usepackage{graphicx}
\usepackage{dcolumn}
\usepackage{bm}
\usepackage{hyperref}
\usepackage{enumitem}
\usepackage{balance}
\usepackage{comment}
\usepackage{cleveref}

\usepackage{amsthm}

\begin{document}
\title{Imaginary time evolution with quantum nondemolition measurements: multi-qubit interactions via measurement nonlinearities}

\author{Manikandan Kondappan\footnote{\href{https://orcid.org/0000-0002-8814-6789}{orcid.org/0000-0002-8814-6789}}}
\affiliation{State Key Laboratory of Precision Spectroscopy, School of Physical and Material Sciences, East China Normal University, Shanghai 200062, China} 
\affiliation{New York University Shanghai, 1555 Century Ave, Pudong, Shanghai 200122, China}

\author{Manish Chaudhary}
\affiliation{State Key Laboratory of Precision Spectroscopy, School of Physical and Material Sciences, East China Normal University, Shanghai 200062, China} 
\affiliation{New York University Shanghai, 1555 Century Ave, Pudong, Shanghai 200122, China}

\author{Ebubechukwu O. Ilo-Okeke} 
\affiliation{New York University Shanghai, 1555 Century Ave, Pudong, Shanghai 200122, China}

\author{Valentin Ivannikov}
\affiliation{New York University Shanghai, 1555 Century Ave, Pudong, Shanghai 200122, China} 
\affiliation{NYU-ECNU Institute of Physics at NYU Shanghai, 3663 Zhongshan Road North, Shanghai 200062, China}

\author{Tim Byrnes}
\email{tim.byrnes@nyu.edu}
\affiliation{New York University Shanghai, 1555 Century Ave, Pudong, Shanghai 200122, China}  
\affiliation{State Key Laboratory of Precision Spectroscopy, School of Physical and Material Sciences, East China Normal University, Shanghai 200062, China}
\affiliation{NYU-ECNU Institute of Physics at NYU Shanghai, 3663 Zhongshan Road North, Shanghai 200062, China}
\affiliation{Center for Quantum and Topological Systems (CQTS), NYUAD Research Institute, New York University Abu Dhabi, UAE}
\affiliation{National Institute of Informatics, 2-1-2 Hitotsubashi, Chiyoda-ku, Tokyo 101-8430, Japan}
\affiliation{Department of Physics, New York University, New York, NY 10003, USA}

\begin{abstract}
We show that quantum nondemolition (QND) measurements can be used to realize measurement-based imaginary time evolution.  In our proposed scheme, repeated weak QND measurements are used to estimate the energy of a given Hamiltonian.  Based on this estimated energy, adaptive unitary operations are applied such that only the targeted energy eigenstates are fixed points of the evolution.  In this way, the system is deterministically driven towards the desired state. The nonlinear nature of the QND measurement, which allows for producing interactions between 
systems, is explicitly derived in terms of measurement operators.  We show that for suitable 
interaction times, single qubit QND Hamiltonians can be converted to effective multi-qubit imaginary time operations. We illustrate our techniques with the example of preparing a four qubit cluster state, which is prepared using only collective single qubit QND measurements and single qubit adaptive operations.  
\end{abstract}

\date{\today}

\maketitle
 
\section{Introduction}

Finding the ground state of a Hamiltonian is an important task that arises in many areas of physics, ranging from condensed matter physics, high energy physics, to quantum chemistry \cite{buluta2009quantum,georgescu2014quantum,aaronson2009quantum,aspuru2012photonic,kokail2019self,banuls2020simulating}. It is also important in the context of general optimization problems, where the task is to find a configuration that minimizes the value of a cost function \cite{papadimitriou1998combinatorial,korte2011combinatorial,mohseni2022ising}. Methods such as quantum annealing and adiabatic quantum computing aim to find the ground state of a complex Hamiltonian by adiabatically changing it from a Hamiltonian with a known ground state \cite{das2008colloquium,hauke2020perspectives,albash2018adiabatic,yarkoni2022quantum}.  An approach that has attracted some interest recently is imaginary time evolution, where the dynamics of a controllable quantum system are made to follow the operator $ e^{-H t} $, where $ H $ is the Hamiltonian of interest.  Such an evolution amplifies the amplitude of the ground state, such that for long times $ t $, the state converges to the ground state.  

While the non-unitary nature of the operator $ e^{-H t} $ means that it is an unphysical process, several approaches have been developed to mimic the dynamics of this operator.  One approach is Variational Imaginary Time Evolution (VITE) \cite{mcardle2019variational} where the parameters of a quantum circuit are adjusted with a hybrid quantum-classical approach to match the imaginary time dynamics \cite{jones2019variational,endo2020variational,yuan2019theory}.  Another approach is the Quantum Imaginary Time Evolution (QITE) method  \cite{motta2020determining}, where the non-unitary time evolution is approximated by a unitary operator.  In Probabilistic Imaginary Time Evolution (PITE) \cite{liu2021probabilistic}, an ancilla qubit is used to approximate the non-unitary evolution, such that the state collapses to the desired state with some probability.  

In a previous paper \cite{mao2022deterministic}, we proposed an alternative approach to perform imaginary time evolution, where the non-unitary dynamics is approximated by a sequence of weak measurements. In the approach, a weak measurement that approximates imaginary time evolution for short times is applied to the system repeatedly. Due to the measurement process being random, a conditional unitary operation is applied dependent upon the measurement outcome, which turns the stochastic process into a deterministic evolution.  The particular measurement that was originally proposed in Ref. \cite{mao2022deterministic} coupled an ancilla qubit to the system, such that a large number of measurements reveals the energy.  If the energy estimate is
higher than the desired value, a unitary operation is applied which acts to create a transition in 
energy space.  In this way, the only stable fixed point of the system is the ground state, which the system approaches exponentially on average, realizing the imaginary time evolution. 

In this paper, we show that quantum nondemolition (QND) measurements can be used to perform the measurements that are required for the imaginary time evolution scheme of Ref. \cite{mao2022deterministic}.   QND measurements are a well-established and experimentally demonstrated technique in numerous contexts \cite{julsgaard2001experimental,chou2005measurement,matsukevich2006entanglement,haas2014entangled,kovachy2015quantum,pu2018experimental,lester2018measurement,zarkeshian2017entanglement,omran2019generation,Hammerer2010,Kuzmich1998,duan2000quantum,kuzmich2004nonsymmetric,duan2000squeezing,moller2008quantum,bao2020spin}. By showing that it is possible to use QND measurements in the imaginary time evolution, this greatly improves the practical realizability of the scheme. 

We also further develop the theory of QND measurements by 
showing how it is able to convert non-interacting single body Hamiltonians to interacting Hamiltonians, using the nonlinearity of quantum measurements.  We show that this nonlinearity can be exploited to realize high order effective interactions. Such high order interactions are useful for generating particular classes of states that have been shown to be useful in a quantum information context. For example, in quantum simulation many interesting Hamiltonians involve three and higher order interactions, such as the toric code Hamiltonian \cite{kitaev2003fault}, stabilizer Hamiltonians for cluster states \cite{raussendorf2001one}, topological many-body localized Hamiltonians,  \cite{PhysRevLett.124.190601,PhysRevA.94.023610,huang2021emulating} and various other quantum error correction algorithms \cite{RevModPhys.87.307}. 
Producing such higher order interactions is often difficult because they do not occur naturally, or they are very small in magnitude. A case in point are cluster (or graph) states, which are the essential resource for one-way quantum computing \cite{clusterstatebriegel,nielsen2006cluster}.  
Even the most basic graph states require a third or higher order interactions \cite{nielsen2006cluster,haselgrove2003quantum,chen2011no}, when prepared as the ground state of a Hamiltonian \cite{bartlett2006simple,van2008graph,kyaw2014measurement,kyaw2018cluster}. This has made their large-scale preparation challenging in a practical setting. 
To illustrate our techniques, we show how our methods can be used to generate a 4-qubit cluster state by imaginary time evolution, which requires a 3-qubit interaction Hamiltonian.  

\section{Outline and Basic Idea of this paper}

Before commencing a more technical discussion of the theory of QND measurements and imaginary time evolution, we give the basic idea of how the measurement-based imaginary time evolution (MITE) technique of Ref. \cite{mao2022deterministic} works.  We also point out the critical results that are derived in this paper, for the benefit of readers who are not interested in the technical details. 

The aim of MITE (as described in Ref. \cite{mao2022deterministic}) is to obtain the ground state of a given Hamiltonian $ H $ in a controllable quantum system.  In the approach, a weak measurement of a quantum state is made in the energy eigenbasis of $ H $, which we call $ | E_n \rangle $.  The initial state can be an arbitrary state, which can be written  
\begin{align}
| \psi_0 \rangle = \sum_n \psi^{(0)}_n  | E_n \rangle .  
\end{align}
In Ref. \cite{mao2022deterministic}, a particular measurement scheme was described, but as we described in this paper, any weak measurement in the energy eigenbasis of the form
\begin{align}
    M(E) \propto \sum_n \exp \left( - \frac{(E_n - E)^2}{2\sigma^2} \right) | E_n \rangle \langle E_n | 
    \label{memeas}
\end{align}
may equally be used.  The above measurement has a Gaussian distribution in energy space, centered at energy $ E $ and has a width $ \sigma $.  For the purposes of this section, we will not concern ourselves with details such as normalization factors required for $ M(E)$ to be a valid quantum measurement for simplicity.  We call $ M(E) $ a weak measurement because it does not cause total collapse in the energy eigenbasis.  In the limit of $ \sigma \rightarrow 0 $, the measurement approaches a strong measurement which results in a complete collapse of the state to a particular energy eigenstate. 

Each measurement outcome $E$ associated with $ M(E) $ occurs randomly, since it is a quantum measurement.  Hence there is no guarantee that we will obtain the desired outcome for the ground state $ E = E_0$. The basic idea of MITE is to cause a ``guided collapse'' such that multiple measurements of $ M(E) $ are made, and adjustments (in the form of unitary operations) are made using the information that is gained from the measurements such as to ensure that the full collapse occurs for  $ E = E_0$. 

Specifically, the procedure proceeds as follows.  First, perform a measurement corresponding to $M(E) $ and obtain an estimate of the energy $E$ where the collapse is occuring.  The state at this point is 
\begin{align}
M(E) |\psi_0 \rangle \propto  \sum_n \psi^{(0)}_n \exp \left( - \frac{(E_n - E)^2}{2\sigma^2} \right) | E_n \rangle . 
\end{align}
The parameters of the Gaussian are chosen such that there is still superposition of the energy states, but the amplitudes are more centered around the energy $E$.  If the readout energy $ E $ is greater than a threshold energy $ E_{\text{th}} $, then this indicates that the collapse is occuring centered at an energy that is higher than desired.  In this case, a unitary operation $ U_C $ is applied, which redistributes the state in the energy eigenbasis.  This has the effect of repopulating the ground state $|E_0 \rangle $ and avoiding convergence on any high energy state.  If the readout energy $ E < E_{\text{th}}$, no unitary is applied, so that the state can continue towards a full collapse towards the ground state $|E_0 \rangle $. 

Full convergence is attained when the same measurement outcome is consistently attained.  Repeated measurements with the same outcome cause a full collapse as may be seen by directly calculating
\begin{align}
    M^k (E_0) \propto &  \sum_n \exp \left( - \frac{k(E_n - E_0)^2}{2\sigma^2} \right) | E_n \rangle \langle E_n | \nonumber \\
   & \overset{k \rightarrow \infty } \longrightarrow |E_0 \rangle \langle E_0 |
\end{align}
The factor of $k $ in the Gaussian reduces its width by a factor of $ \sqrt{k} $.  Once collapse on the energy $ E=E_0 $ is obtained, all higher energy states are strongly suppressed via the Gaussian factor which is an exponential decay factor in energy.  This completes the MITE scheme. 

In Ref. \cite{mao2022deterministic}, an ancilla qubit-based measurement scheme was used to show that it is possible to realize a measurement operator with similar characteristics to (\ref{memeas}).  In this paper, one of the main goals is to show that QND measurements can equally be used to realize a suitable measurement operator to realize the MITE scheme. The POVM for the QND measurement, together with the Gaussian approximations are shown in (\ref{povmdefinition}) and  (\ref{cfuncapprox}).  The QND measurement has some additional features, such as the presence of additional Gaussian peaks with a negative phase as illustrated in Fig. \ref{fig2}.  These require including small modifications of the MITE procedure, such that proper convergence can still be attained.  This is explained in Sec. \ref{sec:qndite}.  While these additional Gaussian peaks may appear to be an undesirable artifact of the QND measurement, these in fact can be taken advantage of to produce multi-body effective interactions.  Conventionally, the QND measurements produce two-body interactions, originating from the Gaussian functional form of the measurement (Sec. \ref{sec:shorttime}).  However, the additional Gaussian peaks can be used to produce third and higher order interactions, as shown in Sec. \ref{sec:qndints}. Finally, we illustrate our methods with an explicit example of generating a four qubit cluster state in Sec. \ref{sec:cluster}.

\section{Generalized quantum nondemolition measurements}
\label{sec:generalized}

We first start by generalizing the theory of QND measurements as developed in Ref. \cite{aristizabal2021quantum, chaudhary2022stroboscopic} to the measurement of an arbitrary Hamiltonian.  This will form the foundation for the measurement-based imaginary time evolution that will be shown in the next section.

\subsection{Wavefunction evolution}

\begin{figure}[t]
\includegraphics[width=\linewidth]{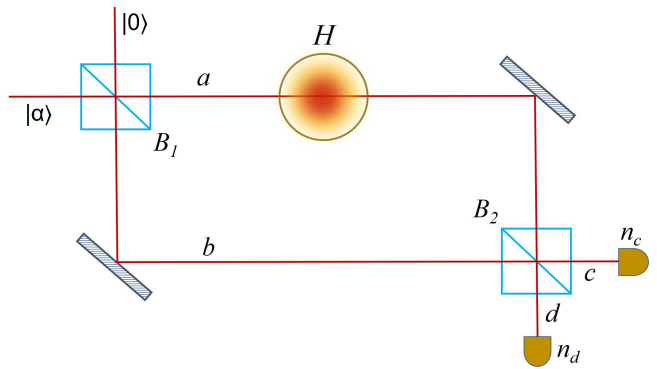}
\caption{Experimental setup for performing a QND measurement. (a) Coherent light $ |\alpha\rangle $ is divided into two modes $a,b$ at beamsplitter $ B_1 $. In this generalized scheme of QND of an arbitrary system, the interaction is only through mode $a$ which interacts with qubit via the QND Hamiltonian as depicted in the figure which is an atom cloud in this case (\ref{qndhamiltonian}).  The modes are then interfered at beamsplitter $B_2$ that are detected using photon counters with outcomes $n_c $ and $ n_d $ respectively. While we show the Mach-Zehnder configuration for conceptual simplicity, any equivalent configuration can also be implemented to realize the QND measurement.  \label{fig1}}
\end{figure}

In a QND measurement, coherent light is arranged in a Mach-Zehnder interferometer and the light on one path interacts with the target system (see Fig. \ref{fig1}).  The form of the interaction is \cite{Hammerer2010,Kuzmich1998}
\begin{align}
H_{\text{int}} = \hbar\Omega H a^\dagger a ,
\label{qndhamiltonian}
\end{align}
where $ a, b $ denote the bosonic annihilation operators of the light in the two arms of the interferometer. 
Here $ H $ is a Hamiltonian that specifies the basis of the QND measurement.  For typical QND measurements, $ H $ is taken to be an operator such as $ \sigma^z $ \cite{aristizabal2021quantum}, where $ \sigma^{x,y,z} $ are Pauli operators.   This Hamiltonian will be eventually that which will be evolved in imaginary time.  We note that while not all QND experiments have the exact implementation as shown in Fig. \ref{fig1}, many can be reduced to this form, and we consider this to be a generic setup for a QND measurement.

The state of the system after interacting with the light is
\begin{align}
| \Psi(t) \rangle & = e^{-i H_{\text{int}} t/\hbar } |\frac{\alpha}{\sqrt{2}} \rangle_a |\frac{\alpha}{\sqrt{2}} \rangle_b | \psi_0 \rangle \nonumber \\
& = \sum_n \psi^{(0)}_n |\frac{\alpha e^{-i E_n \tau }}{\sqrt{2}} \rangle_a |\frac{\alpha}{\sqrt{2}} \rangle_b | E_n \rangle
\end{align}
where $ \tau = \Omega t $ is the dimensionless time and the coherent state for the mode $ a $ is defined as
\begin{align}
| \alpha \rangle_a = e^{- | \alpha |^2/2} e^{\alpha a^\dagger } | 0 \rangle
\end{align}
and similarly for $ b $. Here, $ \alpha $ is the amplitude of the coherent light entering the first beamsplitter in Fig. \ref{fig1}. The initial state of the target system is $ | \psi_0 \rangle  $ which can be expanded in terms of energy eigenstates of $ H | E_n \rangle = E_n | E_n \rangle $
\begin{align}
| \psi_0 \rangle = \sum_n \psi^{(0)}_n  | E_n \rangle ,  
\end{align}
where $ \psi^{(0)}_n = \langle E_n | \psi_0 \rangle $. After interacting with the atoms, the two modes are interfered via the second beam splitter which transforms the modes as 
\begin{align}
c & = \frac{1}{\sqrt{2}} ( a+ b) \nonumber \\
d & = \frac{1}{\sqrt{2}} ( a - b) . 
\end{align}
After the second beam splitter, the photons are detected in the Fock basis. The above sequence modulates the quantum state of the atoms due to the atom-light entanglement that is generated by the QND interaction. The final unnormalized state after detection of $ n_c, n_d $ photons in modes $c,d $ respectively is 
\begin{align}
| \widetilde{\psi}_{n_c n_d}(\tau)\rangle & = \sum_{n} \psi^{(0)}_n   C_{n_c n_d} (E_n \tau)  |E_n \rangle ,
	\label{modfun}
\end{align}
where we defined the function 
\begin{align}
C_{n_c n_d}(\chi) = \frac{\alpha^{n_c+n_d}e^{-|\alpha|^2/2}}{\sqrt{n_c!n_d!}}\cos^{n_c}(\chi)\sin^{n_d}(\chi) . 
\label{modulatingfunc}
\end{align}
The probability of obtaining a photonic measurement outcome $n_c, n_d $ is
\begin{align}
p_{n_c n_d} (\tau) & = \langle  \widetilde{\psi}_{n_c n_d}(\tau)	| \widetilde{\psi}_{n_c n_d}(\tau)\rangle  \nonumber \\
& = \sum_{n} | \psi^{(0)}_n  C_{n_c n_d} (E_n \tau ) |^2 . 
\label{probability}
\end{align}

The function $ C_{n_c n_d}(\chi) $ takes a form of a multi-peak Gaussian, peaked at solutions of the equation
\begin{align}
\cos 2 \chi =  \frac{n_c - n_d}{n_c + n_d} 
\label{itemaxval}
\end{align}
and has a Gaussian width of 
\begin{align}
\sigma_{n_c n_d} \approx \frac{1}{\sqrt{(1+f_{n_c n_d})(n_c+n_d)}} . 
\label{cfuncwidth}
\end{align}
Here we included a factor of 
\begin{align}
f_{n_c n_c} = \frac{4 n_c n_d}{(n_c + n_d)^2}
\end{align}
which has the range $ 0 \le f_{n_c n_c} \le 1 $ and weakly adjusts the Gaussian width. The Gaussian form of the $C$-function causes a collapse of the initial state into one of the energy eigenstates $ |E_n \rangle $ for large $ n_c + n_d \approx |\alpha|^2 $.  For weak coherent light, the state only partially collapses with a modified envelope function as given by $  C_{n_c n_d} (E_n \tau ) $.

\subsection{QND measurement operators}

The effect of performing the QND measurement can be summarized in a simple way: it introduces an additional factor $ C_{n_c n_d} (E_n \tau) $ into the initial wavefunction as given in (\ref{modfun}).  It will be useful to write the QND measurement in terms of a measurement operator, or more precisely a Positive Operator Valued Measure (POVM)  
\begin{align}
M_{n_c n_d} (\tau) & = \sum_{n} C_{n_c n_d}(E_n \tau)  | E_n \rangle \langle E_n  | .
\label{povmdefinition}
\end{align}
The state following the QND measurement (\ref{modfun}) can then be written as
\begin{align}
	| \widetilde{\psi}_{n_c n_d}(\tau)\rangle & = M_{n_c n_d} (\tau) | \psi_0 \rangle  .  
\end{align}
The measurement operators (\ref{povmdefinition}) satisfy the completeness relation
\begin{align}
\sum_{n_c, n_d} M_{n_c n_d} (\tau)^\dagger M_{n_c n_d} (\tau) = I , 
\end{align}
which results from the property of the $ C $-functions 
\begin{align}
\sum_{n_c n_d = 0}^{\infty}|C_{n_c n_d}(\chi)|^2 = 1  . 	
\label{totalprob}
\end{align}

In order to connect the QND measurement with imaginary time evolution, it will be illuminating to approximate the $ C$-function as a Gaussian.  Using Stirling's approximation we obtain the expression \cite{Ilo-Okeke2016}
\begin{align}
C_{n_c n_d} & (\chi) \approx    \nonumber \\
&  
s_{n_c n_d}(\chi) A_{n_c n_d} \exp \left( - \frac{ \left[ |t(\chi)| - \frac{1}{2} \arccos ( \frac{n_c - n_d}{n_c + n_d} ) \right] ^2}{2 \sigma_{n_c n_d}^2 } \right)
\label{cfuncapprox}
\end{align}
where $ t(x)= \arcsin (\sin (x)) $ is the triangular wave and the amplitude is defined as
\begin{align}
A_{n_c n_d} = \left\{
\begin{array}{ll}
\frac{\alpha^{n_c + n_d} e^{- |\alpha|^2/2}}{\sqrt{ (n_c+n_d)!}} & \text { if } n_c n_d = 0 \\
\frac{\alpha^{n_c + n_d} e^{- |\alpha|^2/2}}{\sqrt{ (n_c+n_d)!} 
(\frac{2 \pi n_c n_d}{n_c + n_d})^{1/4}} & \text{ otherwise}
\end{array}
\right.
\end{align}
and the sign of the Gaussian is determined by 
\begin{align}
s_{n_c n_d}(\chi)  = q^{n_c} (\chi + \pi/2) q^{n_d}  (\chi)
\label{sfuncdef}
\end{align}
where $ q(x) = \text{sgn} (\sin(x)) $ is the square wave.  The sign function only takes values $ s_{n_c n_d}(\chi) = \pm 1 $. 

A less accurate, but simpler form of the $ C$-function can be obtained according to 
\begin{align}
C_{n_c n_d} & (\chi) \approx    \nonumber \\
&  
s_{n_c n_d}(\chi) A_{n_c n_d} \exp \left( - \frac{ \left[ \cos(2\chi) -  \frac{n_c - n_d}{n_c + n_d}  \right]^2}{ 8 \tilde{\sigma}_{n_c n_d}^2 } \right)
\label{cfuncapprox2}
\end{align}
where we defined the empirical standard deviation factor
\begin{align}
\tilde{\sigma}_{n_c n_d} = \frac{1}{\sqrt{8 \left( \frac{n_c + n_d}{4} \right)^{
2- f_{n_c n_d} } }} . 
\label{weirdvar} 
\end{align}
The peculiar form of the standard deviation arises due to the fact that at extremal values $ n_c = n_d = 0 $, the dependence within the exponential (\ref{cfuncapprox2}) has a $ \sim \chi^4 $ dependence, rather than a conventional $ \sim \chi^2 $ dependence.  The exponent in (\ref{weirdvar}) adjusts for this such that the width of the peak is of the correct value.  

\begin{figure}[t]
\includegraphics[width=\linewidth]{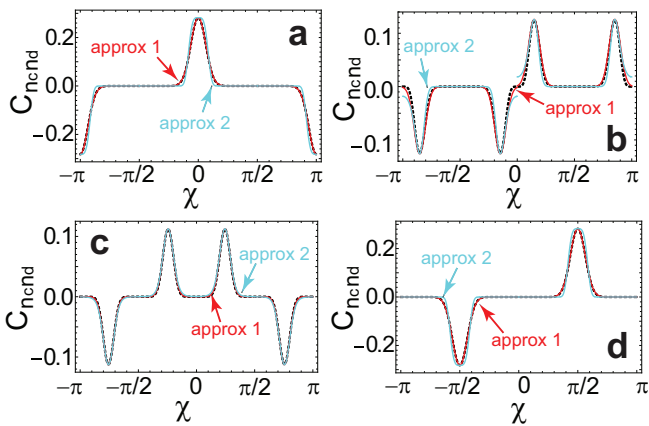}
\caption{Approximating functions to the function $ C_{n_c n_d}(\chi)$.  The two approximate expressions  (\ref{cfuncapprox}) and (\ref{cfuncapprox2}) which we call ``approx 1'' and ``approx 2'' respectively, are shown for the measurement outcomes (a) $n_c = 25, n_d = 0 $; (b) $n_c = 20, n_d = 5 $; (c) $n_c = 13, n_d = 12 $; (d) $n_c = 0, n_d = 25 $. The dotted line is the exact function (\ref{modulatingfunc}) for comparison. We use $ \alpha = 5$ for all plots. \label{fig2}}
\end{figure}

In Fig. \ref{fig2} we show the performance of the two approximations for two parameter values.  We see that both (\ref{cfuncapprox}) and (\ref{cfuncapprox2}) are peaked at the correct values as given in (\ref{itemaxval}), which is guaranteed from the argument of the Gaussian.  The primary difference between (\ref{cfuncapprox}) and (\ref{cfuncapprox2}) is that in the latter approximation, the form of the peak is not of Gaussian form in the region of $ n_c = n_d = 0 $, due to the $ \sim \chi^4 $ dependence in the exponential.  In the cases that we will examine, the precise functional form is not as important as the location of the peaks.  For this reason, later Eq. (\ref{cfuncapprox2}) will be used to approximate the $ C$-function.

\section{Imaginary time evolution}

In this section we show that the QND measurements described in the previous section can be utilized as the imaginary time evolution measurement operators that were introduced in Ref. \cite{mao2022deterministic}.

\subsection{Ancilla qubit based measurement operators}

We first review the imaginary time evolution measurement operators introduced in Ref. \cite{mao2022deterministic}. There we considered the measurement operators
\begin{align}
{\cal M}_l & = \frac{1}{\sqrt{2}} ( \cos \tau H - (-1)^l \sin \tau H ) \\
& =  \frac{1}{\sqrt{2}} \sum_n ( \cos \tau E_n - (-1)^l \sin \tau E_n ) | E_n \rangle \langle E_n |  \nonumber \\
& = \frac{e^{-(-1)^l  H \tau }}{\sqrt{2}} 
\label{itemeasops}
\end{align}
for $ l \in \{0,1 \} $. These measurement operators can be realized by interacting an ancilla qubit with the interaction Hamiltonian $ H \otimes \sigma^y $ for a time $ \tau $ and then measuring the ancilla in the $ \sigma^z $-basis.  After a sequence of measurements, the combined effect can be evaluated to be
\begin{align}
{\cal M}_0^{k_0} {\cal M}_1^{k_1} = \sum_n A_{k_0 k_1 } (\tau E_n) |E_n \rangle \langle E_n |
\label{itemeasurementseq}
\end{align}
where the amplitude function is
\begin{align}
A_{k_0 k_1 } (x) = \cos^{k_0} ( x + \pi/4) \sin^{k_1} (x + \pi/4)  .
\label{amplitudefunc}
\end{align}
The amplitude function is peaked at solutions to the equation
\begin{align}
\sin 2x  =  \frac{k_1 - k_0}{k_0 + k_1} 
\label{itemaxvalold}
\end{align}
and has a Gaussian width of 
\begin{align}
\sigma_{k_0 k_1} \approx \frac{1}{\sqrt{(1+f_{k_0 k_1})(k_0+k_1)}} . 
\end{align}

In the imaginary time evolution approach of Ref. \cite{mao2022deterministic}, Eq. (\ref{itemaxvalold}) is used to estimate the energy readout according to 
\begin{align}
E_{\text{est}} = \frac{1}{2\tau} \arcsin \left(\frac{k_1 - k_0}{k_0 + k_1} \right)  .
\end{align}
In the scheme as described in Ref. \cite{mao2022deterministic}, the energy spectrum was limited to the range $ -\pi/4 \le \tau E_n \le \pi/4 $ such that only the principal value of the arcsine needs to be considered. The imaginary time evolution proceeds as follows. If the energy estimate is higher than a particular threshold $ E_{\text{th}} $ then a corrective unitary $ U_C $ is applied, which induces a transition from the existing state to another energy state.  If the energy estimate is below $ E_{\text{th}} $ then the sequence of measurements is allowed to converge and collapse the state.  In this way, the system's only fixed point is the ground state and the stochastic evolution due to the measurement operators is converted to a deterministic evolution.

\subsection{QND based imaginary time evolution}
\label{sec:qndite}

We now show how the QND measurement operators (\ref{povmdefinition}) can be used in place of the ancilla-based measurement operators for imaginary time evolution.  The basic idea of our approach is to replace the imaginary time measurement operators (\ref{itemeasops}) by a weak QND measurement specified by (\ref{povmdefinition}).  A very weak coherent light source (e.g. $ |\alpha|^2 \sim 1 $) is input to the interferometer of Fig. \ref{fig1}.  This is then repeated a large number of times in a similar way to that given in (\ref{itemeasurementseq}).  In such a situation the QND measurement-induced collapse occurs very slowly, and it is possible to apply a corrective unitary to drive the system to a desired state deterministically in a similar way to the imaginary time scheme as discussed in Ref.\cite{mao2022deterministic}. 

Let us consider a sequence of  $ T $ such QND measurements, and evaluate the effect of the combined measurement.  Using (\ref{povmdefinition}) we find that 
\begin{align}
\prod_{t=1}^T M_{ n_c^{(t)} n_d^{(t)} } (\tau) = \sum_n C_{\text{tot}} (E_n \tau) |E_n \rangle \langle E_n |
\label{cummeas}
\end{align}
where the cumulative measurement function is 
\begin{align}
C_{\text{tot}} (\chi) & = \prod_{t=1}^T C_{  n_c^{(t)} n_d^{(t)} } (\chi) \nonumber \\
& = \frac{\alpha^{n_c^{\text{tot}} + n_d^{\text{tot}}} e^{- T| \alpha |^2/2} }{ 
\prod_{t=1}^T \sqrt{ n_c^{(t)}! n_d^{(t)}! }} \cos^{n_c^{\text{tot}}} (\chi) \sin^{n_d^{\text{tot}}} (\chi)
\label{ctotfunc}
\end{align}
and we defined
\begin{align}
n_c^{\text{tot}} = \sum_{t=1}^T n_c^{(t)} \nonumber \\
n_d^{\text{tot}} = \sum_{t=1}^T n_d^{(t)} . 
\label{ncndtot}
\end{align}

Comparing (\ref{cummeas}) and (\ref{itemeasurementseq}), there is an obvious similarity to the form of the operators.  They are both diagonal in the energy eigenbasis of $ H $, and the functions $ A_{k_0 k_1 } (x) $ and $ C_{\text{tot}} (\chi) $ have a similar form. In terms of the functional dependence they are completely equivalent up to a constant shift $ \chi = x + \pi/4 $.  This means that the measurement readouts can be done with the cumulative photon counts (\ref{ncndtot}) in a similar way to (\ref{itemaxvalold}).  Reusing the results from (\ref{itemaxval}) and (\ref{cfuncwidth}), we see that the function $ C_{\text{tot}} (\chi) $ is peaked at the value
\begin{align}
\cos 2 \chi =  \frac{n_c^{\text{tot}} - n_d^{\text{tot}}}{n_c^{\text{tot}} + n_d^{\text{tot}}} 
\label{itemaxvaltot}
\end{align}
and has a Gaussian width of 
\begin{align}
\sigma_{n_c^{\text{tot}} n_d^{\text{tot}}} \approx \frac{1}{\sqrt{(1+f_{n_c^{\text{tot}} n_d^{\text{tot}}}) (n_c^{\text{tot}} +n_d^{\text{tot}})}} . 
\end{align}
In the case that the energy range is fixed to $ 0 \le \tau E_n \le \pi/2 $, an estimate of the energy can be made from the relation
\begin{align}
E_{\text{est}} (n_c^{\text{tot}}, n_d^{\text{tot}})  = \frac{1}{2\tau} \arccos \left( \frac{n_c^{\text{tot}} - n_d^{\text{tot}}}{n_c^{\text{tot}} + n_d^{\text{tot}}}  \right)  .
\label{eestdef}
\end{align}
We will see that it is also possible to utilize long interaction times where the energy range is larger than $ 0 \le \tau E_n \le \pi/2 $. 

The QND measurement-based imaginary time evolution then proceeds as follows.  After the $(t+1)$th QND measurement is performed, corrective unitaries are applied such that the state becomes
\begin{align}
| \psi_{t+1} \rangle = \frac{U_{ m_c^{(t+1)} m_d^{(t+1)} } V_{n_c n_d}  M_{n_c n_d} | \psi_{t} \rangle}{ 
\sqrt{ \langle \psi_{t} | M_{n_c n_d}^\dagger M_{n_c n_d} | \psi_{t} \rangle }}
\end{align}
where the corrective unitary is
\begin{align}
U_{m_c m_d} = \left\{
\begin{array}{ll}
I & \text{if } | E_{\text{est}} (m_c,m_d) - E_{\text{tgt}} | < \delta_{\text{tgt}} \\
U_C & \text{otherwise}
\end{array}
\right.
\label{uopdef}
\end{align}
and $ V_{n_c n_d} $ is a phase correction unitary in the basis of $ | E_n \rangle  $ that is used to remove the effects of $ s_{n_c n_d} $ as given in (\ref{sfuncdef}).  This will only be important in the long interaction time regime and this will be discussed further in the next section.  The counters are updated as
\begin{widetext}
\begin{align}
m_c^{(t+1)} = \left\{
\begin{array}{ll}
m_c^{(t)} + n_c & \text{if } | E_{\text{est}} (m_c^{(t)} + n_c,m_d^{(t)} + n_d) - E_{\text{tgt}} | < \delta_{\text{tgt}}  \\
0  & \text{otherwise}
\end{array}
\right. \nonumber \\
m_d^{(t+1)} = \left\{
\begin{array}{ll}
m_d^{(t)} + n_d & \text{if } | E_{\text{est}} (m_c^{(t)} + n_c,m_d^{(t)} + n_d) - E_{\text{tgt}} | < \delta_{\text{tgt}} \\
0  & \text{otherwise}
\end{array}
\right. .
\label{counterupdate}
\end{align}
\end{widetext}
In the above we have generalized the condition for applying the corrective unitary slightly from what appeared in Ref. \cite{mao2022deterministic} such that convergence towards a particular target state with energy $ E_{\text{tgt}} $ is obtained.   The $ \delta_{\text{tgt}} $ is a tolerance for the energy estimate being in the vicinity of the target energy.  This modified condition will allow us to target not only the ground state but any state in the energy spectrum.

\section{QND measurement induced effective interactions}
\label{sec:qndints}

One of the uses of QND measurements has been for generating entanglement between spatially separated quantum systems.  For example, in Ref. \cite{julsgaard2001experimental}, entanglement was produced between two atomic ensembles using QND measurements.  One way to understand how entanglement is produced is that initially entanglement is produced between the two atomic ensembles and light in a tripartite fashion, then the light is measured out, swapping the entanglement to between the two atomic ensembles.  In this section, we provide another way of understanding the effective interactions that are produced, making use the measurement operator formalism of Sec. \ref{sec:generalized}.  We first show how effective interactions can be produced using QND measurements in the short interaction time regime, then show how longer interaction times can produce higher order interactions.

\subsection{Short interaction time regime}
\label{sec:shorttime}

In the previous section, we showed that QND measurements could be used as the basis for imaginary time evolution.  In Ref. \cite{mao2022deterministic}, only the short interaction times were considered, which in the QND formulation corresponds to the energy spectrum being in the range $ 0 \le \tau E_n \le \pi/2 $. This range was chosen because the readout of the QND measurement has a multivalued nature and are peaked at the solutions of (\ref{itemaxvaltot}).  Fixing the energy range as above gives a one-to-one relation between the QND measurement readouts and the energies.  For a general Hamiltonian $ H_0 $, this requires adding an energy offset such that the energy spectrum fits in this range. Let us assume that $ H_0 $ has a spectrum that has the ground state and the largest eigenvalue of the same magnitude $ -E_0 = E_{\max} $.  The most sensitive part of the cosine function with respect to the variation of $ \chi $ in (\ref{itemaxvaltot}) is around $ \chi = \pi/4 $, hence we centre the energy spectrum around this point such that
\begin{align}
H \tau = H_0 \tau +\frac{\pi}{4} I .
\label{offsetham0}
\end{align}
The measurement operator (\ref{povmdefinition}) can then be written as
\begin{align}
M_{n_c n_d}(\tau) &  = C_{n_c n_d} (H \tau) \nonumber \\
& \approx \pm A_{n_c n_d} 
\exp \left( - \frac{ \left[ \sin(2 H_0 \tau ) -  \frac{n_d - n_c}{n_c + n_d}  \right]^2}{8 \tilde{\sigma}_{n_c n_d}^2 } \right) , 
\label{mfuncapprox}
\end{align}
where we used the approximate expression (\ref{cfuncapprox2}) and we abbreviated the sign dependence $ s_{n_c n_d} ( H \tau ) $ by writing $ \pm $, since it contributes an unimportant global phase in this case.  For short times $ | E_n \tau| \ll 1 $, we may approximate the sine function linearly, giving
\begin{align}
M_{n_c n_d}(\tau) \approx \pm A_{n_c n_d} 
\exp \left( - \frac{ \left[ H_0 \tau -  \frac{n_d - n_c}{2(n_c + n_d)}  \right]^2}{2 \sigma_{n_c n_d}^2  } \right) , 
\label{mfuncapprox2}
\end{align}
where we replaced the empirical variance factor with the standard variance (\ref{cfuncwidth}) since the argument of the Gaussian no longer has the issue with the fourth power.  

We see from (\ref{mfuncapprox2}) that the effect of QND measurements in the short time regime is to produce a Gaussian with respect to the Hamiltonian $ H_0 $, with an offset related to the measurement readout. This form makes apparent how the imaginary time evolution arises. For measurement outcomes such that $ E_0 \tau > (n_d - n_c)/(2 (n_c+n_d)) $, the Gaussian tail causes exponential damping of the excited states.  The imaginary time evolution as described in the previous section takes advantage of this by using a feedback approach to drive the system towards the desired outcome. 

Let us illustrate the above with a specific example.  
The fact that the measurement operator involves the {\it square} of the Hamiltonian $ H_0 $ is the reason why entanglement generation is possible using QND measurements, even if $H_0 $ itself does not include interaction terms.  Let us show how entanglement generation results using the typical choice for QND measurement
\begin{align}
H_0 = - (\sigma^z_1 + \sigma^z_2) .  
\label{h0exampleham}
\end{align}
Substituting this into (\ref{mfuncapprox2}), we see that the effect of the measurement operator is
\begin{align}
M_{n_c n_d}(\tau) \propto
\exp \left( - \frac{ \sigma^z_1 \sigma^z_2 \tau^2 + 
( \sigma^z_1 + \sigma^z_2) (\frac{n_d - n_c}{2(n_c + n_d)} ) \tau }{\sigma_{n_c n_d}^2  } \right) , 
\label{twoqubitint}
\end{align}
which involves an interaction term originating from the $ H_0^2 $ term in the Gaussian.  

Another way to view this is that the state collapses onto one of the energy eigenstates $ | E_n \rangle $ of $ H_0 $ for large $ n_c + n_d $, where the variance $ \sigma^2_{n_c n_d} \rightarrow 0 $.  The three energy eigenstates of (\ref{h0exampleham}) are $ |00\rangle $, $ c_1 |01\rangle + c_2 |10 \rangle $, and $ |11 \rangle $, where $ c_1, c_2 $ are normalized complex coefficients.  The second of these eigenstates has the form of an entangled state (for $ c_1, c_2 \ne 0 $), resulting in entanglement generation.

\subsection{Long interaction time regime}
\label{sec:longtimeite}

The form of the measurement operator (\ref{mfuncapprox}) suggests there is another way to produce effective interactions by taking advantage of the sine function in the argument of the Gaussian function.  For longer interaction times the contribution of the higher order terms beyond (\ref{mfuncapprox2}) become important, giving rise to further a nonlinearity. To illustrate this, let us consider the example where the Hamiltonian is 
\begin{align}
H_0 = -\sum_{n=1}^N \sigma^z_n 
\label{totalspinham}
\end{align}
which is the total spin of an ensemble and is one that is typically considered in QND measurements.  Evaluating the sine factor in (\ref{mfuncapprox}) for the interaction time $ \tau = \pi/4 $ we have
\begin{align}
\sin \left( \frac{\pi}{2} \sum_{n=1}^N \sigma^z_n  \right) 
& = \cos (\frac{\pi(N+1)}{2}) \prod_{n=1}^N  \sigma^z_n .
\label{productsin}
\end{align}
This shows that it is possible to produce a highly nonlinear effective interaction from a single particle interaction (\ref{totalspinham}). 

\begin{figure}[t]
\includegraphics[width=\linewidth]{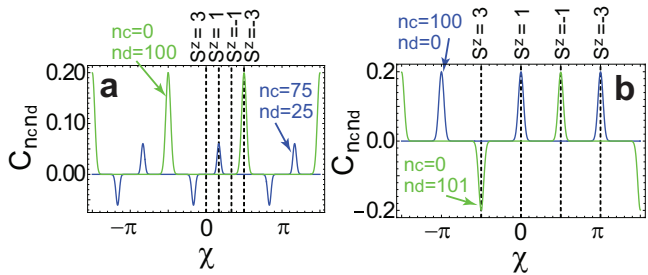}
\caption{Effects of short and long interaction times for QND measurements.  
The $C$-function is plotted for different measurement outcomes as marked in the 
context of the energy levels $ \tau E_n $ marked by the dashed vertical lines. The energy levels
are those given by (\ref{totalspinham}) with $ N = 3 $ with the offset (\ref{offsetham0}).  The interaction times are chosen as (a) $ \tau = \pi/12 $; (b) $ \tau = \pi/4 $. The parameters for (a) satisfy the short time regime as $ 0 \le \tau E_n \le \pi/2 $, while for (b) the interaction times are in the long time regime. We use $ \alpha = 10 $ for all calculations.  \label{fig3}}
\end{figure}

We may take a different point of view to obtain the same result from an energy point of view. In Fig. \ref{fig3} we contrast the short and long interaction time regimes.  In the short interaction time regime (Fig. \ref{fig3}(a)), the energy levels are within a limited range $ 0 \le \tau E_n \le \pi/2 $, hence for a particular $n_c, n_d $ measurement outcome only one of the levels are picked out at a time.  For longer interaction times (Fig. \ref{fig3}(b)), the energy levels are more spaced out, and the multivalued nature of the $ C $-function can pick out more than one energy level at a time.  For example, for the $ n_c = 0, n_d = 101 $ outcome, the $ C$-function picks out states with $ S^z= 3 $ and $ S^z = -1 $, corresponding to the states 
$ |000\rangle $ and $ | 011 \rangle,  | 101 \rangle,  | 110 \rangle $.  These states all satisfy $ \sigma^z_1 \sigma^z_2 \sigma^z_3 = 1 $ and suppress the remaining states, which satisfy $ \sigma^z_1 \sigma^z_2 \sigma^z_3 = -1 $.  This corresponds to an evolution 
\begin{align}
M_{n_c n_d} (\pi/4) \propto s_{n_c n_d} (H \tau) \exp 
\left( \frac{ \left(  \frac{n_d-n_c}{n_c+n_d} \right) \sigma^z_1 \sigma^z_2 \sigma^z_3 }{ 4 \sigma_{n_c n_d}^2}  \right) 
\label{meas3qubit}
\end{align}
according to (\ref{mfuncapprox}). We have reinstated the phase factor $ s_{n_c n_d} (H \tau) $ since in this regime this can contribute a relative phase factor that can affect the state.  For example, in Fig. \ref{fig3}(b), for odd values of $n_d$, there is a relative minus sign between the $ S^z = 3 $ and $ S^z=-1 $ states which can affect the state.

The above is merely a simple example of a way of producing high order interactions by taking advantage of the nonlinearity that is present in (\ref{mfuncapprox}).  By choosing interaction times $ \tau $ that picks out other states in the energy spectrum of the Hamiltonian $ H $, other types of interactions can be generated.

\section{Example: Generation of a four qubit cluster state}
\label{sec:cluster}

We now illustrate our QND measurement-based imaginary time evolution methods for the case of generating a four-qubit cluster state. The four-qubit cluster state is defined as,
\begin{align}
| C_4 \rangle = \frac{1}{2} \left( 
|0000\rangle+|0011\rangle+|1101\rangle+|1110\rangle \right) .  
\label{clusterc4}
\end{align}
This example serves as not only an illustrative example which shows how QND measurements can be used to generate Hamiltonians involving beyond-second order interactions but is also a practically important example as it is an essential component for gate teleportation of a CNOT gate \cite{gottesman1999demonstrating}.

\subsection{Stabilizer Hamiltonian}

A useful way to define cluster states is in terms of stabilizers, which are operators that form a group such that the cluster state is their simultaneous eigenstate with eigenvalue $ +1 $ \cite{nielsen2006cluster, fowler2008topological}.  In the case of (\ref{clusterc4}), the stabilizers are
\begin{align}
{\cal S} = \{ \sigma^z_1 \sigma^z_2, \sigma^x_3 \sigma^x_4, \sigma^x_1 \sigma^x_2 \sigma^x_3, \sigma^z_2 \sigma^z_3 \sigma^z_4 \} .  
\label{stablizersc4}
\end{align}
The stabilizers can be used to construct a Hamiltonian that has the cluster state as its ground state, simply by summing the stabilizer elements
\begin{align}
H_{C_4} = - \sigma^z_1 \sigma^z_2 - \sigma^x_3 \sigma^x_4  - \sigma^x_1 \sigma^x_2 \sigma^x_3- \sigma^z_2 \sigma^z_3 \sigma^z_4 .  
\end{align}
This can be used in the context of imaginary time evolution, by starting from an arbitrary initial state $ | \psi_0 \rangle $ the system is driven 
\begin{align}
e^{- H_{C_4} t} | \psi_0 \rangle  & = e^{\sigma^z_2 \sigma^z_3 \sigma^z_4 t } e^{\sigma^x_1 \sigma^x_2 \sigma^x_3 t } e^{\sigma^x_3 \sigma^x_4 t } e^{\sigma^z_1 \sigma^z_2 t }  | \psi_0 \rangle  \nonumber \\
& \overset{t \rightarrow \infty}{\longrightarrow} |C_4 \rangle 
\label{c4ite}
\end{align}
where we used the fact that all stabilizers in (\ref{stablizersc4}) mutually commute.  
Hence by applying a sequence of imaginary time evolutions, the four qubit cluster state can be prepared.

\subsection{Measurement and correction operators}

We now construct the measurement and correction operators required to realize the imaginary time evolution in (\ref{c4ite}).  We see that there are two basic types of operators involved, the two qubit interaction $ \sigma^z_1 \sigma^z_2, \sigma^x_3 \sigma^x_4 $ and three qubit interaction $ \sigma^x_1 \sigma^x_2 \sigma^x_3, \sigma^z_2 \sigma^z_3 \sigma^z_4 $.  We have already seen that such two and three qubit operators are possible to realize in Sec. \ref{sec:qndints}.  With some small modifications of what we have already discussed, we will show how these can be implemented.

\subsubsection{Two qubit interactions}
\label{sec:twoqubitints}

First let us consider the two qubit imaginary time operator $ e^{\sigma^z_1 \sigma^z_2 t } $. As discussed in Sec. \ref{sec:shorttime}, this can be achieved using a short time evolution and taking advantage of the Gaussian form of the $ C$-function.  While the measurement operator (\ref{twoqubitint}) has the correct basic form, it has the wrong sign on the $ \sigma^z_1 \sigma^z_2 $ term.  This can be rectified by instead choosing the Hamiltonian 
\begin{align}
H = \sigma^z_1 - \sigma^z_2 .  
\label{twoqubitqndham}
\end{align}
The three eigenstates of (\ref{twoqubitqndham}) are 
\begin{align}
|E_0 \rangle & = |10 \rangle \nonumber \\
|E_1 \rangle & = c_1 |00 \rangle + c_2 |11 \rangle \nonumber \\
|E_2 \rangle & = |01 \rangle
\end{align}
where $ c_1, c_2 $ are normalized complex coefficients, and the energies are
\begin{align}
E_0 & =- 2  \nonumber \\
E_1 & = 0 \nonumber \\
E_2  & =  2  .  
\end{align}
The state we would like to target is the state $  c_1 |00 \rangle + c_2 |11 \rangle $ which is the ground state of the Hamiltonian $  - \sigma^z_1 \sigma^z_2 $.  Hence we target the energy 
\begin{align}
E_{\text{tgt}} = E_1 = 0,
\end{align}
which corresponds to the measurement outcome $ n_c>0 , n_d = 0 $ according to (\ref{eestdef}).  We choose a time in the short time regime such that the remaining energy levels are not solutions of the location of the peaks given by (\ref{itemaxvaltot}).  That is, for $ E_n = \pm 2 $, we choose a time such that 
\begin{align}
\cos 2 E_{\text{tgt}} \tau \ne \cos 2 E_n \tau .  
\end{align}
This avoids picking up additional states when targeting $ | E_1 \rangle $.  An example of an interaction time that satisfies this is $ \tau = \pi/8 $.   

Finally, we require an operator that induces a transition between non-target energy states and the target state $ | \langle E_{\text{tgt}}  | U_C |  E_n \rangle | > 0 , \forall n $.  From inspection of the energy eigenstates we may choose
\begin{align}
U_C = \sigma^x_1 . 
\end{align}
The phase correction unitary does not need to be considered for this case since $ s_{n_c n_d} (H\tau) $ only contributes a global phase, and we can take
\begin{align}
V_{n_c n_d} = I . 
\end{align}
With this, we may then follow the procedure described in Sec. \ref{sec:qndite}, where convergence to the state $  c_1 |00 \rangle + c_2 |11 \rangle $ is attained. 

For the operator $ e^{\sigma^x_3 \sigma^x_4 t } $, the procedure is the same up to a basis change.  Specifically, 
the Hamiltonian for the QND measurement is 
\begin{align}
H = \sigma^x_3 - \sigma^x_4 .  
\label{twoqubitqndhamx}
\end{align}
while the correction operator is 
\begin{align}
U_C = \sigma^z_1 .  
\end{align}

\subsubsection{Three qubit interactions}
\label{sec:threequbitints}

We next show how to perform the imaginary time evolution corresponding to $ e^{\sigma^x_1 \sigma^x_2 \sigma^x_3 t } $.  The basic approach was already described in Sec. \ref{sec:longtimeite}, where a $ N $-qubit interaction in the $ z $ basis was realized.  In our case, we choose the Hamiltonian 
\begin{align}
H = - (\sigma^x_1 + \sigma^x_2  + \sigma^x_3) + 3 I
\label{3qubith0}
\end{align}
and we choose an interaction time in the long time regime such a three qubit interaction similar to (\ref{meas3qubit}) is generated. The eigenstates of (\ref{3qubith0}) are
\begin{align}
|E_0 \rangle & = |+++ \rangle \nonumber \\
|E_1 \rangle & = c_1 |++ - \rangle + c_2 |+-+ \rangle + c_3 |-++ \rangle \nonumber \\
| E_2 \rangle & = c_1 |+- - \rangle + c_2 |-+- \rangle + c_3 |+-- \rangle \nonumber \\
|E_3 \rangle & = | --- \rangle
\end{align}
where $ c_1, c_2, c_3 $ are normalized complex coefficients, and the energies are
\begin{align}
E_0  & =  0 \nonumber \\
E_1 & = 2 \nonumber \\
E_2 & = 4\nonumber \\
E_3  & = 6 .  
\end{align}
The ground states of the desired Hamiltonian $ - \sigma^x_1 \sigma^x_2 \sigma^x_3 $ are $ |+++ \rangle, |+- - \rangle,  |-+- \rangle,  |+-- \rangle $, which correspond to the combination of the $ |E_0 \rangle $ and $ |E_2 \rangle $ states.  To simultaneously target both of these states, we require a time such that a particular outcome $ n_c, n_d $ has solutions at multiple values of (\ref{itemaxval}).  Namely, we require
\begin{align}
\cos 2 E_0 \tau & = \cos 2 E_2 \tau 
\end{align}
but
\begin{align}
\cos 2 E_0 \tau & \ne \cos 2 E_1 \tau \nonumber \\
\cos 2 E_0 \tau & \ne \cos 2 E_3 \tau
\end{align}
which can be satisfied with $ \tau = \pi/4 $.  Then from (\ref{eestdef}), we set the target state to be 
\begin{align}
E_{\text{tgt}} = E_0 = 0 , 
\end{align}
which also will target $ |E_2 \rangle $. 

When we work in the long time regime where multiple energy levels are picked out, we must also handle the energy dependent prefactor $ s_{n_c n_d} (H \tau) $ that appears in (\ref{meas3qubit}).  Depending on the parity of $ n_c, n_d $, this adds a relative $(-1)$ phase factor between the $ |E_0 \rangle $ and $|E_2 \rangle $ states, as can be seen from the $ n_c= 0, n_d = 101 $ curve in Fig. \ref{fig3}(b). To eliminate this phase, after performing each three qubit measurement we may apply the conditional unitary
\begin{align}
V_{n_c n_d} & = e^{i ( (n_c + n_d) \text{mod} 2) H \tau } ,
\label{conditionalunitary}
\end{align}
where $ H $ is given by (\ref{3qubith0}).  We note that this only involves single qubit rotations which we assume are readily available.

The correction operator should induce a transition between the states $ \{ |E_0 \rangle, |E_2 \rangle \} $ and $ \{ |E_1 \rangle, |E_3 \rangle \} $ hence we choose
\begin{align}
U_C = \sigma^z_1
\end{align}
and otherwise follow the procedure in  Sec. \ref{sec:qndite}.  

For the $ e^{\sigma^z_2 \sigma^z_3 \sigma^z_4  t } $ operator, the procedure is the same up to a basis change.  For example, 
the Hamiltonian for the QND measurement is 
\begin{align}
H = - ( \sigma^z_2  + \sigma^z_3+ \sigma^z_4) + 3 I
\label{3qubith0z}
\end{align}
and the correction operator is
\begin{align}
U_C = \sigma^x_4 .
\end{align}
The conditional operator is the same as (\ref{conditionalunitary}) with the Hamiltonian (\ref{3qubith0z}).

\subsection{Numerical evolution}

We now use the operators defined in the previous section to perform the measurement-based imaginary time evolution. Our simulation proceeds in four stages, one for each of the imaginary time operators in (\ref{c4ite}).  Starting from a random initial state $ | \psi_0 \rangle $, we first perform the two two-qubit operations following the procedure in Sec. \ref{sec:qndite} with the operators as defined in Sec. \ref{sec:twoqubitints}.  This is then followed by the two three-qubit operations following the procedure in Sec. \ref{sec:qndite} with the operators as defined in Sec. \ref{sec:threequbitints}. The measurement outcomes are chosen according to Born probabilities, see the Appendix of Ref. \cite{PhysRevA.106.033314} for the procedure.  

In order to assess the success of the procedure, we evaluate the fidelity of the state during the evolution with respect to the cluster state (\ref{clusterc4}) defined as
\begin{align}
F_{C} = | \langle C_4 | \psi \rangle |^2 .  
\label{overallfid}
\end{align}
We also evaluate whether the state has been projected into the correct subspace using 
\begin{align}
F_1 & = \langle \psi | \frac{1}{2} \left( I + \sigma^z_1 \sigma^z_2 \right) |\psi \rangle  \nonumber \\
F_2 & = \langle \psi | \frac{1}{2} \left( I + \sigma^x_3 \sigma^x_4 \right) |\psi \rangle  \nonumber \\
F_3 & = \langle \psi | \frac{1}{2} \left( I + \sigma^x_1 \sigma^x_2 \sigma^x_3 \right) |\psi \rangle  \nonumber \\
F_4 & = \langle \psi | \frac{1}{2} \left( I + \sigma^z_2 \sigma^z_3 \sigma^z_4\right) |\psi \rangle .
\label{sectorfids}
\end{align}

\begin{figure}[t]
\includegraphics[width=\linewidth]{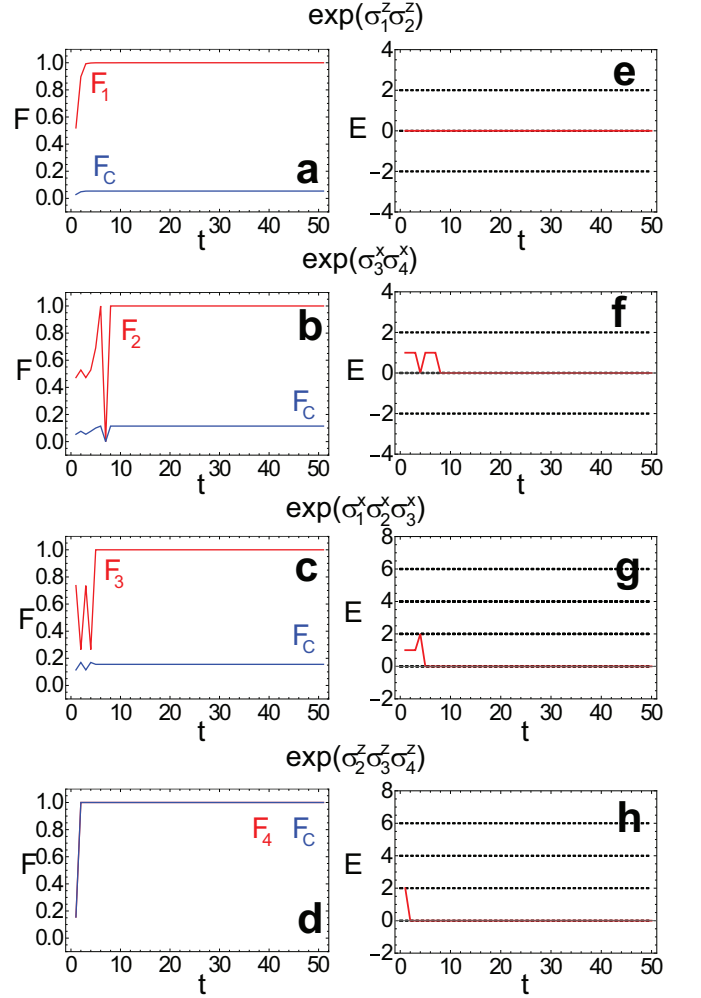}
\caption{Preparing a four qubit cluster state with imaginary time evolution.  (a)-(d) show fidelities as defined in (\ref{overallfid}) and (\ref{sectorfids}) as marked.  (e)-(h) show the energy estimates (\ref{eestdef}) based on the cumulative measurement outcomes (\ref{counterupdate}).  The procedure starts from a random initial state and applies the imaginary time evolution corresponding to (a)(e) $ e^{\sigma^z_1 \sigma^z_2 t } $, (b)(f) $ e^{\sigma^x_3 \sigma^x_4 t }  $, (c)(g) $  e^{\sigma^x_1 \sigma^x_2 \sigma^x_3 t }  $, (d)(h) $ e^{\sigma^z_2 \sigma^z_3 \sigma^z_4 t } $ in sequence.  We use parameters $ \alpha = 1 $, $ \delta_{\text{tgt}} =0.5 $ throughout.  Numerically we impose a photonic cutoff of 5 photons.  
\label{fig4}}
\end{figure}

Our results are shown in Fig. \ref{fig4}. We show an example of a single run of the algorithm.   From the fidelity plots Figs. \ref{fig4}(a)-(d) we see that within 10 rounds of measurement and unitary correction, the state settles into a steady state, where further measurement rounds do not affect the state.  The associated energy estimates in  Figs. \ref{fig4}(e)-(h) show that the measurements have converged to the target energy, which is $ E_{\text{tgt}} = 0 $ for all cases.  The fidelities for each subspace (\ref{sectorfids}) in each case attain a value of unity at steady state.  The fidelity with respect to the cluster state however remains fairly low until the last step, where unit fidelity is attained.  Due to the stochastic nature of the imaginary time algorithm, each run produces a different trajectory, but in all cases, the final state is the cluster state (\ref{c4ite}), which is attained with unit fidelity.

\section{Summary and conclusions}
\label{sec:conc}

We have shown that QND measurements can be used as the measurement required in the imaginary time evolution scheme as proposed in Ref. \cite{mao2022deterministic}.  This is possible because the photon number readouts of the QND measurement can be used to estimate the energy of the QND Hamiltonian.  Then using an adaptive unitary operation based on the energy readout, where the system is disturbed unless the energy is in the desired range, the system deterministically converges to a target state.  Our scheme of Sec. \ref{sec:qndite} is a  modified version of the one introduced in Ref. \cite{mao2022deterministic}, where an arbitrary state of the energy spectrum 
can be targeted, not only the ground state.  By considering long interaction times, we also showed how multiple states in the energy spectrum can be targeted.  In the case of a QND Hamiltonian that is the total spin of $N $ qubits, this can generate a $ N $th order effective interaction.  We illustrated the technique by the deterministic preparation of a four qubit cluster state, which  requires two and three qubit interactions in terms of the ground state of a Hamiltonian. Of particular note is that the cluster state could be generated without any interactions of the qubits, only collective measurements of their total spin.  Namely, the QND Hamiltonians $ H $, the correction operators $ U_C $, and the phase correction operator $ V_{n_c n_d} $ are all single qubit Hamiltonians.  The nonlinearity arises purely from the measurement.  The cluster state example is merely an example, and other types of quantum states can be targetted in a similar way to that demonstrated in Ref. \cite{mao2022deterministic}. 

There are several attractive aspects of the scheme that we discuss in this paper.  First, QND measurements are a well-known 
method of performing quantum measurements, and implemented in several different platforms \cite{julsgaard2001experimental,chou2005measurement,matsukevich2006entanglement,haas2014entangled,kovachy2015quantum,pu2018experimental,lester2018measurement,zarkeshian2017entanglement,omran2019generation,Hammerer2010,Kuzmich1998,duan2000quantum,kuzmich2004nonsymmetric,duan2000squeezing,moller2008quantum,bao2020spin}.   In Ref. \cite{mao2022deterministic}, while a concrete scheme for the measurements
was proposed, it was based on an interaction with an ancilla qubit which is not as easily translated into an experimental setting.  The QND scheme that we propose has a very flexible geometry that allows physical qubits to be measured with two completely independent laser beams.  The Mach-Zehnder configuration allows for highly separated qubits to be entangled even when the line-of-sight is obstructed. Second, high order interactions of arbitrary order can be easily generated simply by adding more spins within the Mach-Zehnder interferometer.  Assuming that the laser beams can be redirected to illuminate various qubits at will, this allows for a way of creating entanglement between any combination of the qubits, including multi-qubit interactions.  This allows for an unlimited range of multipartite entanglement to be generated \cite{walter2016multipartite,miyake2003classification,dur2000three,radhakrishnan2020multipartite}, simply by reconfiguring the beams in a suitable way.   We note that although our presentation has been limited to the pure state case for simplicity, the measurement-based imaginary time evolution is equally applicable for mixed states. In this case, the formalism of Sec. \ref{sec:qndite} would be adapted to the mixed state case, while the discussion of Sec. \ref{sec:generalized} and Sec. \ref{sec:qndints} would be unchanged as the POVMs are equally valid for mixed states. 

The main challenges of the scheme are primarily in the demands of the QND measurement. QND measurements can introduce decoherence to the atomic states via photon loss and spontaneous emission in producing the QND Hamiltonian, which produces a dephasing effect on the target system \cite{gao2022decoherence}. Photon loss is a less serious effect than spontaneous emission in this situation, due to the use of coherent states of light, which are resilient under loss.  The dephasing must therefore be controlled by working with a sufficiently large detuning to realize the QND Hamiltonian. For the imaginary time scheme that is discussed in this paper, first, an accurate estimate of the energy must be performed according to (\ref{eestdef}), which relies upon an accurate photodetection. Imperfections such as detector inefficiencies and dark counts can affect the energy estimate, degrading the performance of the scheme.  However,  in the short time regime, 
the energy estimate (\ref{eestdef}) is not extremely sensitive to detector inefficiencies, since it is a ratio of the total photon counts, which cancel for large photon numbers. Hence we believe that the short time regime to produce two qubit interactions will be relatively robust experimentally. 
However, in the long time regime, the phase correction unitary can depend upon the parity of $n_c + n_d $, as seen in  (\ref{conditionalunitary}), which requires near perfect photon detection efficiency.  In addition to the higher decoherence rates for long QND interaction times \cite{gao2022decoherence}, this makes the long interaction time schemes more challenging experimentally.  Another challenge is reaching the long time regime as typically in QND measurements the interaction is realized by second order interactions, which can be rather weak.  Thus reaching the long time regime may require additional modifications to the free space setup as implied in Fig. \ref{fig1}, such as the use of cavities.  

We note that generating larger scale cluster states should not be much more difficult than the four qubit example that we showed here.  All stabilizers commute and would involve further sequences of imaginary time operators, in a similar way to (\ref{c4ite}).  For a general cluster state, up to five qubit interactions would be required, which can be generated using the long time regime as shown in (\ref{productsin}). The very fast convergence of each of the imaginary time evolutions as shown in Fig. \ref{fig4} suggests that the routine would be highly scalable, and each additional vertex adds a constant overhead.  As already shown in Ref. \cite{mao2022deterministic}, more complex Hamiltonian ground states can also be realized, although this may involve more complex QND Hamiltonians that involve interactions.  Such a framework is useful for other alternative models of quantum computing such as spinor quantum computing \cite{byrnes2012macroscopic,Abdelrahman2014} where QND measurements are the primary way that entanglement is generated between the ensembles.

 \bibliographystyle{apsrev}
 \bibliography{paperrefs}

\end{document}